\documentclass[1p,sort&compress]{elsarticle} 
\journal{Annals of Physics}
\usepackage[utf8]{inputenc} 
\usepackage{amsmath}
\usepackage{graphicx} 
\usepackage{bbm}
\usepackage{bm}
\usepackage{amssymb}
\usepackage{color}
\usepackage[english]{babel}
\usepackage[pdftex,colorlinks]{hyperref}

\hypersetup{urlcolor=blue}
\def\U{\ensuremath{\mathcal{U}}}
\def\P{\ensuremath{\mathcal{P}}}

\def\id{\ensuremath{\mathbbm{1}}}
\DeclareMathOperator{\Tr}{Tr}
\begin{document}
\begin{frontmatter}
\title{Weak values and weak coupling maximizing the output of weak measurements}
\author{Antonio {Di Lorenzo}}
\ead{dilorenzo@infis.ufu.br}
\address{Universidade Federal de Uberl\^{a}ndia, Uberl\^{a}ndia, MG, Brazil}
\address{CNR-IMM-UOS Catania (Universit\`a), Consiglio Nazionale delle Ricerche,
Via Santa Sofia 64, 95123 Catania, Italy}
\begin{abstract}
In a weak measurement, the average output $\langle o\rangle$ of a probe that measures an observable $\Hat{A}$ of a quantum system undergoing both a preparation in a state $\rho_\mathrm{i}$ and a postselection in a state $E_\mathrm{f}$ is, to a good approximation, a function of the weak value $A_w=\Tr[E_\mathrm{f}\Hat{A}\rho_\mathrm{i}]/\Tr[E_\mathrm{f}\rho_\mathrm{i}]$, a complex number. 
For a fixed coupling $\lambda$, when the overlap $\Tr[E_\mathrm{f}\rho_\mathrm{i}]$ is very small, $A_w$ diverges, 
but $\langle o\rangle$ stays finite, often tending to zero for symmetry reasons. 
This paper answers the questions: what is the weak value that maximizes the output for a fixed coupling? 
what is the coupling that maximizes the output for a fixed weak value? 
We derive equations for the optimal values of $A_w$ and $\lambda$, and provide the solutions. 
The results are independent of the dimensionality of the system, and they apply to a probe having a Hilbert space of arbitrary dimension. Using the Schr\"{o}dinger-Robertson uncertainty relation, we demonstrate that, in an important case, the amplification $\langle o\rangle$ cannot exceed the initial uncertainty $\sigma_o$
in the observable $\Hat{o}$, we provide an upper limit for the more general case, and a strategy to 
obtain $\langle o\rangle\gg \sigma_o$.  
\end{abstract}
\end{frontmatter}


\section{Introduction}
In 1988 \cite{Aharonov1988}, Aharonov et al. introduced the concept of postselected weak measurements, initiating 
a prolific avenue of research. Recently, there have been several works considering 
the possibility of using weak and intermediate strength interaction in order 
to reconstruct an unknown quantum state \cite{Hofmann2010,Lundeen2011,Lundeen2012,Fischbach2012,Salvail2013,DiLorenzo2013a,Wu2013,DiLorenzo2013f} 
and to diminish the noise in a variable by preceding its measurement with the observation of the conjugate variable  \cite{DiLorenzo2013c}. 
Other works, instead, in line with the initial proposal of Aharonov et al. \cite{Aharonov1988}, have focused 
on the amplification effect of weak measurement \cite{Ritchie1991,Hosten2008,Dixon2009,Brunner2010,Gorodetski2012}.
(However, the actual advantage over techniques based on strong measurement  
has been questioned \cite{Knee2013a,Tanaka2013,Knee2013b,Ferrie2013}). 
Indeed, the provocative title of the original paper by Aharonov et al. was ``How the result of a measurement of a component of the spin of a spin-1/2 particle can turn out to be 100'', meaning that the average output of a detector 
which, in a strong measurement, would give $p=\pm\lambda/2$ as outputs, can be amplified to $\langle p\rangle\sim 100\lambda$ 
if the measurement is weak and the system is suitably postselected. The weakness of the measurement means 
that initially the uncertainty over the pointer variable is much larger than $\lambda$, the distance between the peaks expected 
in the strong regime. 
This means that initially the pointer may not be in the zero position, but it could read, e.g., $p_0=99\lambda$ or $p_0=-101\lambda$, etc.  
However, if the detector were a classical object, these fluctuations would cancel out on the average. 
Instead, as the detector obeys quantum mechanics as well, if it is prepared in a suitably quantum coherent superposition of pointer states \cite{Peres1989,Duck1989} and if the system is postselected appropriately, this cancellation does not occur, possibly leading to a large output. 

How large can the average output be? According to the simple formula of Ref.~\cite{Aharonov1988} there are no bounds 
to the average output, but as it turns out, the formula breaks down when the output is largish. 
This has prompted the need to provide a more reliable formula, working also in the regime where 
the measurement strength is weak but the overlap between the preparation and the postselection is small \cite{DiLorenzo2008,Wu2011,DiLorenzo2012a,Kofman2012}.  

For a spin 1/2, it is possible to work out an exact solution for an instantaneous interaction \cite{Peres1989,Duck1989} 
and more generally for a nondemolition interaction of finite duration \cite{DiLorenzo2008}. 
This has allowed to study the maximization of the output based on the exact expression in Ref.~\cite{DiLorenzo2008}, 
and then in Refs.~\cite{Koike2011,Zhu2011,Nakamura2012} with varying degrees of generality. 
Kofman \emph{et al.} \cite{Kofman2012} 
considered some particular cases of the maximization for a detector with an infinite dimensional 
Hilbert space performing a canonical von Neumann measurement, i.e. using a position variable $\Hat{q}$ to couple with the system and its conjugate variable $\Hat{p}$ as the readout for the measurement. 
Furthermore, Ref.~\cite{Kofman2012} only considered 
pure preparation and postselection, so that there may be, in principle, higher 
maxima or lower minima for mixed preparation and postselection. 
During the completion of the present manuscript, a preprint 
appeared \cite{Pang2013} that treats the problem of the maximization of the output as well. 
 The variational approach of Ref.~\cite{Pang2013}, however, does not allow, to determine for which values $A_w$ the maximization is attained, and it applies only when $\Hat{q}$ has a continuous spectrum and the readout $\Hat{p}$ is its conjugate variable. 
To the best of my knowledge, there is no systematic study of the maximization of an arbitrary output variable $\langle o\rangle$ for higher-dimensional systems. The main result of this paper is provided in Eqs.~\eqref{eq:result},\eqref{eq:sol},\eqref{eq:tradeoff}, and \eqref{eq:sol2}.

%
\section{Background}
\subsection{Measurement model.}
As customary when treating weak measurements, it is supposed that a detector interacts with the measured system through the 
Hamiltonian $H=-\lambda \delta(t) \Hat{q} \Hat{A}$, i.e. the von Neumann model \cite{vonNeumann1932} of measurement is assumed.  
In this model, the output variable is usually taken to be $\Hat{p}$, the conjugate variable of $\Hat{q}$, i.e. $[\Hat{q},\Hat{p}]=i$, with $[,]$ the commutator. Thus, $\Hat{q}$ is assumed to have a continuous unbounded spectrum, so that it 
can be treated as a position operator. 
In the following, however, we shall not make this assumption, and in this sense we are diverging from the von Neumann model. 
Instead, we shall consider the output variable $\Hat{o}$ to be arbitrary. Thus, the detector could have a finite-dimensional Hilbert space, for instance it could be a spin 1/2, with $\Hat{q}$  a spin component, etc.  


Before the interaction, the measured system is prepared in a state $\rho_\mathrm{i}$ and the detector in a state $\rho_\mathrm{det}$, so that the total state is 
\begin{equation}
\rho=\rho_\mathrm{i}\otimes\rho_\mathrm{det}.
\label{eq:instate}
\end{equation} 
For simplicity, in the following we shall consider $\rho_\mathrm{i}$ and $\rho_\mathrm{det}$ to be given at time $t=0^-$, immediately before the interaction (otherwise, one should trivially propagate the states forward in time with the non-interacting Hamiltonian). 
The joint state after the interaction is thus
\begin{equation}
\rho^+=
\exp(i\lambda\Hat{A}\Hat{q}) (\rho_\mathrm{i}\otimes\rho_\mathrm{det}) \exp(-i\lambda\Hat{A}\Hat{q}).
\label{eq:rhoaft}
\end{equation}
After the system has interacted with the detector, the latter is observed, usually determining the value of $p$. 
The system, on the other hand, is supposed to undergo 
another measurement yielding an output $F$, 
to which a nonnegative operator $E_F$ is associated.  
When the value of this measurement coincides with some arbitrarily fixed value $F=f$, the output $p$ of the detector is selected 
and analyzed separately. This procedure is known as postselection. 

\subsection{Final state and output.}
The conditional state of the detector, given that the system is successfully postselected in the state $E_\mathrm{f}$, is then 
\begin{equation}
\rho_{\mathrm{det}|f}=
N^{-1}
\Tr_\mathrm{sys}[(E_\mathrm{f}\otimes\id)\rho^+]
,
\label{eq:rhocond}
\end{equation}
with the normalization being but the probability of successful postselection
\begin{equation}
N=\P(E_\mathrm{f})= \Tr_{\mathrm{sys,det}}[(E_\mathrm{f}\otimes\id)\rho^+].
\label{eq:ppost}
\end{equation}
The average value of an observable of the detector, $\Hat{o}$, conditioned on the postselection $f$ is thus 
\begin{equation}
\langle o\rangle =\Tr_\mathrm{det}[\Hat{o}\rho_{\mathrm{det}|f}]=
\frac{M}{N},
\label{eq:genp}
\end{equation}
with 
\begin{equation}
M=\Tr_\mathrm{sys,det}[(E_\mathrm{f}\otimes\Hat{o})\rho^+].
\label{eq:unnormp}
\end{equation}

\subsection{Approximations.}
In the weak measurement limit, the propagator is expanded up to first order, 
\begin{equation} 
\exp(i\lambda\Hat{A}\Hat{q})\simeq 1+i\lambda \Hat{A}\Hat{q} .
\label{eq:wappr}
\end{equation} 
This approximation, applied to Eq.~\eqref{eq:unnormp}, gives 
\begin{equation}	
M\simeq M_1=  \overline{\Hat{o}}\omega
+i\lambda \overline{\Hat{o}\Hat{q}}\alpha 
-i\lambda \overline{\Hat{q}\Hat{o}} \alpha^*
+\lambda^2 \overline{\Hat{q}\Hat{o}\Hat{q}} \beta,
\label{eq:unnormp1}
\end{equation}
where we introduced 
\begin{subequations}
\begin{align}
\omega&= \Tr_\mathrm{sys}[E_\mathrm{f}\rho_\mathrm{i}],\\
\alpha&= \Tr_\mathrm{sys}[E_\mathrm{f}\Hat{A}\rho_\mathrm{i}],\\  
\beta&= \Tr_\mathrm{sys}[E_\mathrm{f}\Hat{A}\rho_\mathrm{i}\Hat{A}], 
\end{align}
\label{eq:weaknorm}
\end{subequations}
and denoted the averages with respect to the initial state of the detector as 
\begin{equation}
\overline{\Hat{O}}=\Tr_\mathrm{det}[\Hat{O}\rho_\mathrm{det}].
\end{equation} 
The probability of postselection, on the other hand, has the expansion 
\begin{equation}
N\simeq N_1=\omega
+i\lambda \overline{\Hat{q}}(\alpha-\alpha^*) 
+\lambda^2  \overline{\Hat{q}^2}\beta.
\label{eq:ppost1}
\end{equation}
We note that for nearly orthogonal preparation and postselection, both $\omega$ and $\alpha$ tend to zero, the former 
faster than the latter, while $\beta$ stays finite (we exclude the trivial cases where $\Hat{A}$ has its eigenstates coinciding 
with those of either $E_\mathrm{f}$ or $\rho_\mathrm{i}$). Hence, in order to give meaningful expressions for all possible preparations 
and postselections, one must retain the second-order terms. Notice that we are departing from the na\"{\i}ve Taylor 
expansion, which prescribes that the propagator be expanded up to second order for consistency \cite{DiLorenzo2012jj}. 
See the Appendix for a further discussion.   	 

The approximation \eqref{eq:wappr} is not to be taken as an operator equation, since $\Hat{q}$ may have an 
unbounded spectrum; instead, Eq.~\eqref{eq:wappr} must be interpreted as meaning that, when it is plugged into Eqs.~\eqref{eq:ppost} and \eqref{eq:unnormp}, it yields a good approximation, provided that $\rho(q,q')$ vanishes 
sufficiently fast for large $q$. 
More precisely, we may give a sufficient condition: if 
\begin{equation}
(2\lambda)^n \max\{|A|\}^n\overline{\Hat{q}^{2n}}^{1/2}\le \delta^n , \forall n\in\mathbb{N}
\label{eq:cond}
\end{equation} 
with $\delta$ a small positive number, then we may apply Eq.~\eqref{eq:wappr}, yielding a 
discrepancy between the actual value and the approximate value within 
$\varepsilon=\Tr[E_\mathrm{f}](e^\delta-1-\delta)$, $|N-N_1|<\varepsilon$, 
while the difference between the actual value of Eq.~\eqref{eq:unnormp} 
and the approximate value is $|M-M_1|\lesssim (1+u)\varepsilon\overline{\Hat{o}^2}^{1/2}$, under some conjecture.\footnote{In general, the approximation can not hold for any operator $\Hat{o}$. For instance, in Ref.~\cite{DiLorenzo2012a}, we have proved that the approximation breaks down for $\Hat{o}=\Hat{p}^n$, if $n$ is sufficiently large. See the Appendix.} 

Reference \cite{Aharonov1988} considers the canonical von Neumann measurement, with $\Hat{o}=\Hat{p}$ the conjugated variable of $\Hat{q}$, and, in addition to the hypothesis \eqref{eq:wappr}, it makes another assumption, 
namely that it is possible expand $N^{-1}$ in a Taylor 
series in $\lambda$, 
\begin{equation}
N^{-1}\simeq \omega^{-1}\left[1+2\lambda\overline{\Hat{q}} A''_w-\lambda^2\left(\overline{\Hat{q}^2} B_w
-4\overline{\Hat{q}}^2 A''^2_w\right)\right],
\label{eq:linappr}
\end{equation}
with $A_w=\alpha/\omega$ the canonical weak value and  $B_w=\beta/\omega$ a positive real number, the second 
weak value. 
For brevity, we defined $A'_w=\mathrm{Re}(A_w)$, $A''_w=\mathrm{Im}(A_w)$.  
We may call this further assumption the polynomial approximation. 
The conditions for its validity are more clearcut, since ordinary numbers are involved, 
not operators: 
\begin{align}
\lambda\overline{\Hat{q}}A''_w&\ll 1, &
\lambda^2\overline{\Hat{q}^2} B_w\ll 1.
\label{eq:hyp2}
\end{align} 
Aharonov et al. then assume that the resulting expansion for $\langle p\rangle$ can be truncated to first order, 
and they also consider a Gaussian state for the detector, which implies that $\overline{\Hat{q}\Hat{p}+\Hat{p}\Hat{q}}-2\overline{\Hat{q}}\, \overline{\Hat{p}}=0$, 
leading to the formula 
$\langle p\rangle \simeq \lambda \mathrm{Re}(A_w)$.

As discussed elsewhere \cite{DiLorenzo2008,Wu2011,DiLorenzo2012a,DiLorenzo2012jj}, 
one cannot always expand $N^{-1}$ in a Taylor 
series in $\lambda$, since to lowest order $N\simeq \Tr_\mathrm{sys}[E_\mathrm{f}\rho_\mathrm{i}]$ and when the preparation and the postselection 
are nearly orthogonal $\Tr_\mathrm{sys}[E_\mathrm{f}\rho_\mathrm{i}]\simeq 0$. 
Here, we shall not make the assumption \eqref{eq:hyp2}, as we are allowing $A_w$ and $B_w$ to take
all possible values. 
Thus, we shall use the interpolating formula derived in Ref.~\cite{DiLorenzo2012a} 
\begin{equation}
\langle o\rangle \simeq \frac{\overline{\Hat{o}}+\lambda 
\left( -i\overline{[\Hat{q},\Hat{o}]} A'_w-\overline{\{\Hat{q}, \Hat{o}\}}A''_w\right)+\lambda^2 \overline{\Hat{q} \Hat{o}\Hat{q}} B_w}{1-2\lambda \overline{\Hat{q}}A''_w+\lambda^2 \overline{\Hat{q}^2} B_w},
\label{eq:start}
\end{equation}

We recall that $B_w\ge |A_w|^2$. The equality holds whenever the postselection $E_\mathrm{f}$ and the preparation $\rho_\mathrm{i}$
have, respectively, eigenstates $|f\rangle$ and $|i\rangle$ with nonzero eigenvalues such that 
$\langle f|\Hat{A}|i\rangle/\langle f|i\rangle = constant$ for all $|f\rangle$ and all $|i\rangle$, with the convention that 
if $\langle f|i\rangle=0$ then also $\langle f|\Hat{A}|i\rangle=0$. See the Appendix for a proof of this statement, which 
was provided, without demonstration, in Refs.~\cite{DiLorenzo2012a,DiLorenzo2012e}. 
In particular, if both $E_\mathrm{f}$ and $\rho_\mathrm{i}$ represent pure states, then $f$ and $i$ can take only a single value, thus 
$B_w=|A_w|^2$. 

%
\section{Statement of the problem.}
Equation~\eqref{eq:start} can be simplified through the transformations $\Hat{q}\to\Hat{q}-\overline{\Hat{q}}=\delta\Hat{q}$, 
$\rho_\mathrm{i}\to \exp(i\lambda \overline{\Hat{q}}\Hat{A})\rho_\mathrm{i}\exp(-i\lambda \overline{\Hat{q}}\Hat{A})$ so that without loss 
of generality we can replace $\delta\Hat{q}\to \Hat{q}$, $\overline{\delta\Hat{q}}=0$. We also put $\delta \Hat{o}=\Hat{o}-\overline{\Hat{o}}$, 
and  we reabsorb the coupling constant $\lambda$ by redefining 
$\Hat{A}\to\lambda \sigma_q\Hat{A}$, with $\sigma_q=\overline{\delta \Hat{q}^2}^{(1/2)}$, so that 
\begin{equation}
\langle \delta o\rangle \simeq \frac{-i\overline{[\Hat{\xi},\delta\Hat{o}]}A'_w-\overline{\{\Hat{\xi},\delta \Hat{o} \}}A''_w+\overline{\Hat{\xi}\delta \Hat{o}\Hat{\xi}} B_w}{1+  B_w},
\label{eq:start2}
\end{equation}
with $\Hat{\xi}=\delta\Hat{q}/\sigma_q$ a normalized variable having zero mean and unit variance. 
Equation \eqref{eq:start2} is our starting point. The goal of this paper is to find the extrema of $\langle \delta o\rangle$.
Since, in general,  $B_w\ge |A_w|^2$, then, in our extremal problem, 
the domain of the variables is the volume bound by the paraboloid of rotation 
$B_w=A'^2_w+A''^2_w$ that contains the point $(A'_w=0,A''_w=0,B_w=1)$. 
Notice that, since the coupling constant $\lambda$ was reabsorbed in the rescaling of $A_w$ and $B_w$, the extrema thus found 
will not depend on it.

We consider a fixed preparation of the probe (otherwise, it can be proved that there is no bound to the average output \cite{Susa2012,DiLorenzo2013h}), and we look for an extremal value of $\langle o\rangle-\overline{\Hat{o}}$ as a function of 
the preparation $\rho_\mathrm{i}$ and the postselection $E_\mathrm{f}$. In a $d$-dimensional Hilbert space, these states are 
characterized by a total of $2d^2-2$ real parameters (we remind the readers that $E_\mathrm{f}$ need not have trace one, but the weak values are invariant upon rescaling of $E_\mathrm{f}$, so that $E_\mathrm{f}$ is effectively characterized by $d^2-1$ parameters, as $\rho_\mathrm{i}$). 
Some of these parameters are superfluous. For instance, if we consider the unitary transformations that leave the observable 
$\Hat{A}$ invariant, $U \Hat{A} U^\dagger =\Hat{A}$, by changing $E_\mathrm{f}$ to $U^\dagger E_\mathrm{f} U$ and $\rho_\mathrm{i}$ to  
$U^\dagger \rho_\mathrm{i} U$, the weak values remain the same. 
In the simplest case ($d=2$) of $\Hat{A}$ representing a spin 1/2, 
the family $U$ is characterized by the rotations around the direction of $\Hat{A}$, so that it is a one-parameter symmetry. 
The total number of real parameters is thus $5$, and it grows with higher dimension $d$ (we discard the trivial case $\Hat{A}=1$).
However, all these parameters enter Eq.~\eqref{eq:start} only through the three real 
combinations $A'_w$, $A''_w$, and $B_w$. 
We recall that, in order to find the extrema of a function $f(t_1,\dots,t_n)=g[x_1(t_1,\dots,t_n),\dots,x_m(t_1,\dots,t_n)]$, 
with $m< n$, 
one just needs to find the extrema of $g(x_1,\dots,x_m)$, simplifying the problem to the maximization of a function of less
variables.

%
\section{Solution.} 
\begin{figure}
\includegraphics[width=4in]{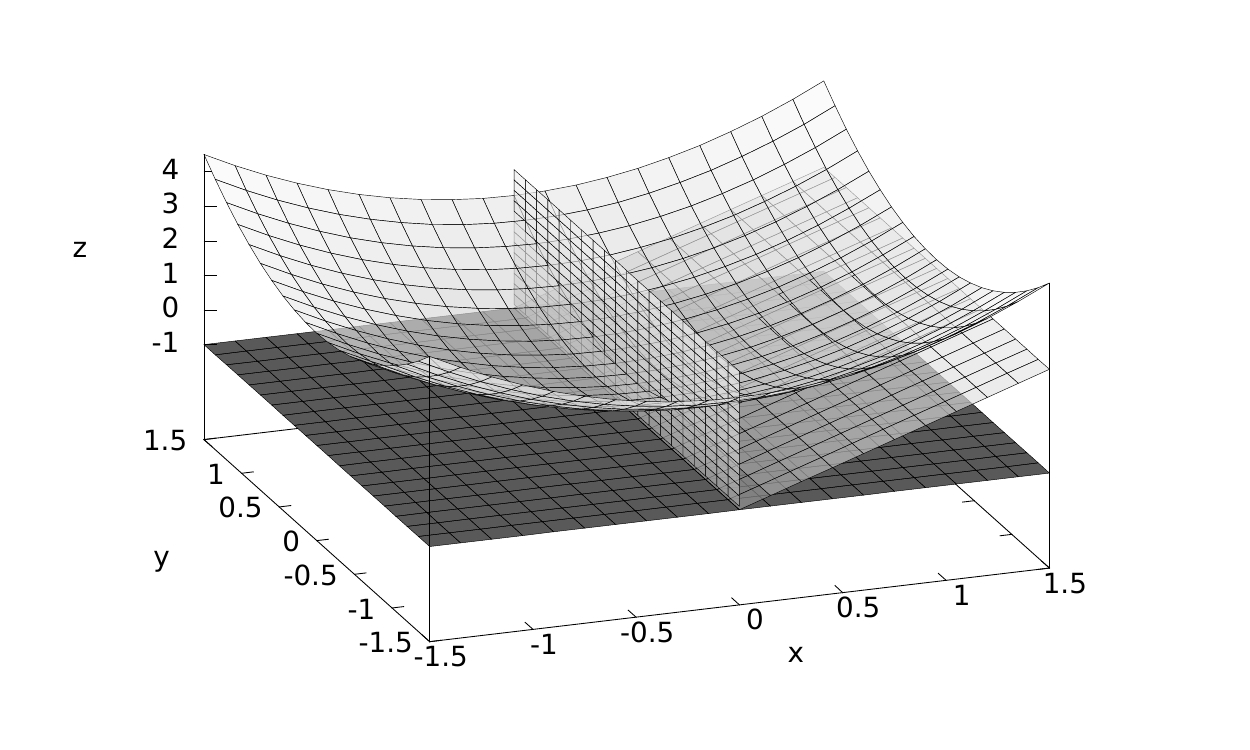}
\caption{\label{fig}An illustration of three of the planes given in Eq.~\eqref{eq:planes} and the paraboloid of rotation~\eqref{eq:boundary}. We put $c=1$, $a=s=0$. The horizontal plane corresponds to $\langle \delta o\rangle\to \infty$, 
the vertical one to $\langle \delta o\rangle=0$, and the oblique one is the tangent plane, obtained for $\langle \delta o\rangle=\langle \delta o\rangle_\mathrm{max}=c/2$.}
\end{figure}

For brevity, we define three Hermitian operators on the Hilbert space of the detector: the anticommutator, the commutator,   
and the sandwich, 
\begin{subequations}
\begin{align}
\Hat{a}=&\ \{\delta\Hat{o},\Hat{\xi}\},\\
\Hat{c}=&\ i[\delta\Hat{o},\Hat{\xi}],\\
\Hat{s}=&\ \Hat{\xi}\delta\Hat{o}\Hat{\xi},
\end{align}
\end{subequations}
and we define their initial averages as $a=\overline{\Hat{a}}$, $c=\overline{\Hat{c}}$, $s=\overline{\Hat{s}}$. 
We assume that at least one among $a$, $c$, and $s$ is non-null, otherwise $\langle \delta o\rangle$ is identically zero within 
the approximation considered.
In order to work with a familiar notation, we define $A'_w=x$, $A''_w=y$, $B_w=z$. 
Thus Eq.~\eqref{eq:start2} reads 
\begin{equation}
c x-a y+(s -\langle \delta o\rangle) z =
\langle \delta o\rangle .
\label{eq:planes}
\end{equation}
Equation \eqref{eq:planes} represents a one-parameter family of planes in the space $\mathbb{R}^3$, the parameter 
being $\langle \delta o\rangle$. 
The problem consists in finding the maximum and minimum value of $\langle \delta o\rangle$ for which the planes intersect 
the allowed region
\begin{equation}
\mathcal{R}:= \{(x,y,z):z\ge x^2+y^2 \} .
\label{eq:region}
\end{equation}
We denote the boundary of $\mathcal{R}$ by	
\begin{equation}
\partial \mathcal{R}:= \{(x,y,z):z= x^2+y^2 \} .
\label{eq:boundary}
\end{equation}
Let us consider the limit $\langle \delta o\rangle\to \pm \infty$. The plane in Eq.~\eqref{eq:planes} tends to 
$z=-1$, and it does not intersect the region $\mathcal{R}$. 
On the other hand, for $\langle \delta o\rangle=s$, the plane is described by the equation $cx-ay=s$, so that it is 
parallel to the $z$-axis, and hence it certainly intercepts the paraboloid \eqref{eq:boundary}  
(the case $a=c=0$ shall be treated separately). 
See Figure \ref{fig} for an illustration. 
Thus, when $\langle \delta o\rangle$ increases from $s$ 
to $+\infty$, there must be a maximum value $\langle \delta o\rangle_\mathrm{max}$ such that for 
$\langle \delta o\rangle<\langle \delta o\rangle_\mathrm{max}$ the plane \eqref{eq:planes} intercepts 
the paraboloid, while for $\langle \delta o\rangle>\langle \delta o\rangle_\mathrm{max}$ it does not. 
As the planes vary continuously with the parameter $\langle \delta o\rangle$, 
the plane with $\langle \delta o\rangle=\langle \delta o\rangle_\mathrm{max}$ must be tangent to the paraboloid. 
Analogously, there exists a minimum value $-\infty<\langle \delta o\rangle_\mathrm{min}<s$ so that the plane 
with $\langle \delta o\rangle=\langle \delta o\rangle_\mathrm{min}$ is also a tangent plane of the paraboloid. 

We recall that the equation for the plane tangent to the surface having implicit equation $\Phi(x,y,z)=0$ in the point 
$P_0=(x_0,y_0,z_0)$ is 
\begin{equation}
\left.\frac{\partial \Phi}{\partial x}\right|_{P_0}\!\!\!\!\!\!(x-x_0)+\left.\frac{\partial \Phi}{\partial y}\right|_{P_0}\!\!\!\!\!\!(y-y_0)
+\left.\frac{\partial \Phi}{\partial z}\right|_{P_0}\!\!\!\!\!\!(z-z_0) =0.
\end{equation}
For the paraboloid $\partial\mathcal{R}$ \eqref{eq:boundary}, 
\begin{equation}
2 x_0 x + 2 y_0 y -z = x_0^2+y_0^2 ,
\end{equation}
where we used $z_0=x_0^2+y_0^2$. 
Thus, comparing the equation 
for the plane tangent to the paraboloid $\partial\mathcal{R}$ \eqref{eq:boundary} to the equation 
\eqref{eq:planes}, we have
\begin{subequations}
\begin{align}
2kx_0=&\,c,
\label{eq:multa}
\\
-2ky_0=&\,a,
\label{eq:multb}
\\
-k=&\,s-\langle \delta o\rangle
\label{eq:multc}
,\\
k(x^2_0+y_0^2) =&\, 
 \langle \delta o\rangle ,
\label{eq:multd}
\end{align}
\end{subequations}
with $k$ a real constant. 
We use Eqs.~\eqref{eq:multa} and \eqref{eq:multb} to find $x_0=c/2k$ and $y_0=-a/2k$, then substitute in Eq.~\eqref{eq:multd},  and 
multiply the result by Eq.~\eqref{eq:multc} to eliminate $k$, finding thus 
\begin{equation}
 \langle \delta o\rangle (s-\langle \delta o\rangle)=\frac{c^2+a^2}{4},
\end{equation}
yielding the two extremal values 
\begin{equation}
\langle \delta o\rangle_\mathrm{m}=\ \frac{1}{2}\left(
s\pm\sqrt{c^2+a^2+s^2}\right).
\label{eq:result}
\end{equation}
Finally, we find $k$ by substituting the solution~\eqref{eq:result} in Eq.~\eqref{eq:multc}, which, substituted into 
Eqs.~\eqref{eq:multa} and \eqref{eq:multb} yields  
\begin{subequations}
\begin{align}
x_0=&\frac{c}{c^2+a^2}
[s\pm \sqrt{c^2+a^2+s^2}]
 ,\\
y_0=&\frac{-a}{c^2+a^2}[s\pm \sqrt{c^2+a^2+s^2}]
 .
\end{align}
\label{eq:sol}
\end{subequations}
Equations~\eqref{eq:sol} provide the location of the maximum and the minimum. Thus, an experimentalist wishing to maximize 
the output, could use Eq.~\eqref{eq:sol} to establish the desired optimal weak value $A^\mathrm{opt}_w=x_0+iy_0$, and then design the experiment with $\rho_\mathrm{i}$ and $E_\mathrm{f}$ such that $A_w=A^\mathrm{opt}_w$. 

We remark that our general results \eqref{eq:result} and \eqref{eq:sol} reduce to those of Ref.~\cite{Kofman2012} 
for $s=0$ and for $\Hat{o}=\Hat{p}$, which implies that 
$c=i\overline{[\Hat{p},\Hat{q}]}=1$. Furthermore, Ref.~\cite{Kofman2012} assumed the case of pure preparation and postselection, so that $B_w=|A_w|^2$ by assumption, and did not demonstrate that the absolute extrema of 
$\langle\delta o\rangle$ are actually reached on the surface $\partial \mathcal{R}$. 
Furthermore, \eqref{eq:result} provides a tighter bound than what was estimated in Ref.~\cite{Kofman2012}.

We recall the Schr\"{o}dinger-Robertson uncertainty relation \cite{Schrodinger1930,Robertson1930}, 
binding the product of the variances of any two variables $\Hat{U}$ and $\Hat{V}$,  
$4\sigma_U^2 \sigma_V^2\ge \left(\overline{-i[\Hat{U},\Hat{V}]}\right)^2+\left(\overline{\{\Hat{U},\Hat{V}\}}\right)^2$, which in our case is $4\sigma_o^2\ge c^2+a^2$, 
We note that the latter inequality implies 
a tradeoff relation 
\begin{equation}
|\langle \delta o\rangle| \le 
\left|\overline{\Hat{O}}\right|+\sqrt{\sigma_o^2+\left(\overline{\Hat{O}}\right)^2},
\label{eq:tradeoff}
\end{equation}
with $\Hat{O}=\Hat{s}/2$. 
In particular, it often happens that $\overline{\Hat{O}}=0$. In this case, Eq.~\eqref{eq:tradeoff} 
simplifies to 
\begin{equation}
|\langle \delta o\rangle| \le 
\sigma_o,
\label{eq:tradeoff2}
\end{equation}
i.e., the maximum shift cannot exceed the initial spread $\sigma_o$. 
Thus, if one wants the average output to exceed the bound $\sigma_o$, one needs $\overline{\Hat{O}}\neq0$, and 
the following strategy should be adopted: 
\\
(\emph{i}) Consider the density matrices for the detector $\tilde{\rho}_\mathrm{det}$ that have a kernel containing 
the kernel $K$ of the operator $\Hat{\xi}$, i.e. $\forall |\psi\rangle$ such that $\Hat{\xi}|\psi\rangle = 0$, then 
$\tilde{\rho}_\mathrm{det} |\psi\rangle=0$. 
In practice, this means that $\tilde{\rho}_\mathrm{det}$ is block-diagonal, with the block in the kernel $K$ being zero, the block acting on the orthogonal complement of $K$ being nonzero, and all other off-diagonal blocks being zero as well.\\
(\emph{ii}) Choose one of these matrices such that $\Tr_\mathrm{det}[\Hat{o}\tilde{\rho}_\mathrm{det}]$ is large.\\
(\emph{iii}) Prepare the detector in an initial state $\rho_\mathrm{det}=
\Hat{\xi}^{-1}_C\tilde{\rho}_{\mathrm{det}|C}\Hat{\xi}^{-1}_C\oplus \rho_{\mathrm{det}|K}$, 
with $\rho_{\mathrm{det}|K}$ a positive operator restricted to the subspace $K$, the kernel of $\Hat{\xi}$, 
while $\tilde{\rho}_{\mathrm{det}|C}$ is the restriction of $\tilde{\rho}_\mathrm{det}$ to the subspace $C$, the orthogonal complement of $K$. Note that $\Hat{\xi}^{-1}$ is well defined in the subspace $C$. Furthermore, if the spectrum $S$ of $\Hat{\xi}$ 
is continuous in a neighborhood of 0, the prescription is to restrict the search to the $\tilde{\rho}_\mathrm{det}$ 
for which $\lim\limits_{\varepsilon\to 0} \int_{S\backslash [-\varepsilon,\varepsilon]} d\xi \xi^{-2} \langle \xi|\tilde{\rho}_\mathrm{det}|\xi\rangle$ is finite.\\
(\emph{iv}) Choose $\rho_{\mathrm{det}|K}$  so that $\Tr_\mathrm{det}[\rho_\mathrm{det}]=1$ and $\Tr_\mathrm{det}[\Hat{o}\rho_\mathrm{det}]=0$. 
\\
Then, if condition (\emph{iv}) can be satisfied, by construction $s=\Tr_\mathrm{det}[\Hat{o}\Hat{\xi}{\rho}_\mathrm{det}\Hat{\xi}]=\Tr_\mathrm{det}[\Hat{o}\tilde{\rho}_\mathrm{det}]$ is as large as one wishes. 

%
\subsection{Optimizing the coupling constant.} One could also be interested in choosing the optimal coupling constant $\lambda$ for 
fixed preparation and postselection of the system. 
Now the problem is simpler, as one has only one real variable. 
We note that it is no longer legitimate to gauge out $\overline{\Hat{q}}$, as the weak values are fixed. 
Hence, we shall find the extrema of 
\begin{equation}
\langle \delta o\rangle = \frac{(c A_w'-a A_w'')\lambda + s B_w \lambda^2}{1-2\xi A''_w \lambda + (1+\xi^2) B_w\lambda^2}
,
\label{eq:restart}
\end{equation}
considered as a function of $\lambda$, where $\xi=\overline{\Hat{\xi}}=\overline{\Hat{q}}/\sigma_q$ and $\lambda$ was restored through the position $\Hat{A}\to \lambda \Hat{A}$. 
A straightforward calculation yields 
\begin{align}
\lambda_\mathrm{m}=&\frac{s\pm \sqrt{s^2+4(c A_w'-a A_w'')[(1+\xi^2)(c A_w'-a A_w'')+2s\xi A''_w]/B_w}}{(1+\xi^2)(c A_w'-a A_w'')+2s\xi A''_w}.
\label{eq:sol2}
\end{align}
The extremal values of $\langle\delta o\rangle$ are obtained by substituting Eq.~\eqref{eq:sol2} into Eq.~\eqref{eq:restart}, 
and they are too complicated to write down here. 
Furthermore, one should be careful not to use the above equation if it yields values of $\lambda$ so large that the 
perturbative expansion \eqref{eq:wappr} breaks down. 
%
\section{Discussion}
We have provided a general framework to find the extremal values of a weak measurement. 
The approach used here has a geometric interpretation in terms of a family of planes in the three-dimensional parameter space defined by the complex weak value $A_w=x+i y$ and by the real weak value $B_w=z$. 
In addition to deriving what are the extremal values, we have provided the more important information, what is their location, 
which can be achieved by an appropriate choice of preparation and postselection. We have also discussed a strategy 
to achieve a maximization going beyond the limit of the Schr\"{o}dinger-Robertson relation. Finally, we have solved the related 
problem of choosing an optimal coupling constant.

%
\section*{Acknowledgments}
This work was performed as part of the Brazilian Instituto Nacional de Ci\^{e}ncia e
Tecnologia para a Informa\c{c}\~{a}o Qu\^{a}ntica (INCT--IQ) and 
it was supported by the Conselho Nacional de Desenvolvimento Cient\'{\i}fico e Tecnol\'{o}gico (CNPq) 
through Process no. 245952/2012-8. 
%
\appendix
\renewcommand*{\thesection}{\Alph{section}}
\section{Modified Taylor expansion}
The joint probability of observing the total system in the final state $E_\mathrm{f}\otimes \Pi_o$ (we assume that the measurement on the detector is sharp, so that Born's rule applies) is 
\begin{equation}
\P(o,E_\mathrm{f})=\Tr{\left[(E_\mathrm{f}\otimes \Pi_o)\U( \rho_\mathrm{i}\otimes \rho_\mathrm{det}) \U^\dagger\right]},
\label{app:joint}
\end{equation}
with the time-evolution $\U=\exp{[i\lambda\Hat{A}\Hat{q}]}$ and the projector $\Pi_o=|o\rangle\langle o|$. We assume the interaction in the von Neumann protocol, 
with $\Hat{A}$ the observable of the system being measured, $\Hat{q}$ an observable of the meter, and $\lambda$ a coupling constant. However, notice that, contrary to the von Neumann protocol, we are not assuming that the readout variable $\Hat{o}$ 
is conjugated to $\Hat{q}$, nor that the meter is initially in a sharp state of the readout 
$\rho_\mathrm{det}\simeq |o=0\rangle\langle o=0|$. 

The probability of postselecting the system in $E_\mathrm{f}$ is 
\begin{equation}
\P(E_\mathrm{f})=\sum_o \P(o,E_\mathrm{f}) .
\label{app:marg}
\end{equation}
Let us apply perturbation theory to Eqs.~\eqref{app:joint} and \eqref{app:marg}, including up to first order terms in the propagator
\begin{align}
\P(o,E_\mathrm{f})\simeq& \Tr\left\{(E_\mathrm{f}\otimes\Pi_o) 
\left[1+i\lambda \Hat{A} \Hat{q} \right]
(\rho_\mathrm{i}\otimes\rho_\mathrm{det}) 
\left[1-i\lambda \Hat{A} \Hat{q}\right]\right\},\\
\P(E_\mathrm{f})\simeq& \Tr\left\{(E_\mathrm{f}\otimes \id) 
\left[1+i\lambda \Hat{A} \Hat{q} \right]
(\rho_\mathrm{i}\otimes\rho_\mathrm{det}) 
\left[1-i\lambda \Hat{A} \Hat{q} \right]\right\}.
\end{align}
The textbook calculus approach would be, e.g., to retain the first order terms, 
\begin{align}
\P_1(o,E_\mathrm{f})=& \omega \overline{\Pi_o}
 +\left\{ i\lambda \alpha  \overline{\Pi_o\Hat{q}} +c.c.\right\} ,\\
\P_1(E_\mathrm{f})=& \omega +\left\{ i\lambda \alpha \overline{\Hat{q}}+c.c.\right\}.
\end{align}
However, the above expressions do not preserve the positivity of the probability, since they are not of the form 
$\P=\Tr{[E U_1 F U_1^\dagger]}$ with $E,F$ positive operators and $U_1$ an arbitrary operator. 
True, the neglected terms are $\mathcal{O}(\lambda^2)$, but nevertheless the probability could turn negative for some values 
of $E_\mathrm{f}$ and $o$ if to lowest order $\P\ll 1$. This can occur if $\lambda$ is somewhat largish. 
The correct way to make the expansion is to keep the product of the first order terms in $\U$ and $\U^\dagger$, giving 
\begin{align}
\P(o,E_\mathrm{f})\simeq&\ \omega 
\overline{\Pi_o}+\lambda \left\{ i\alpha \overline{\Pi_o\Hat{q}} +c.c.\right\}
+
\lambda^2 \beta\overline{\Hat{q}\Pi_o\Hat{q}} ,
\label{joint0}
\\
\P(E_\mathrm{f})\simeq&\ \omega +\left\{ i\lambda \alpha \overline{\Hat{q}} +c.c.\right\}+
\lambda^2 \beta \overline{\Hat{q}^2}.
\label{marg0}
\end{align}
where the system enters the probabilities with three terms: 
the overlap $\omega=\Tr_\mathrm{sys}[E_\mathrm{f}\rho_\mathrm{i}]$ (a positive real number), the complex number $\alpha=\Tr_\mathrm{sys}[E_\mathrm{f}\Hat{A}\rho_\mathrm{i}]$ and the positive real number $\beta=\Tr_\mathrm{sys}[E_\mathrm{f}\Hat{A}\rho_\mathrm{i}\Hat{A}]$. Notice, however, that these probabilities do not sum up to one, but 
$\sum_{E_\mathrm{f}} \P(E_\mathrm{f}) = \sum_{o,E_\mathrm{f}} \P(o,E_\mathrm{f}) = 1+\lambda^2 \langle \Hat{A}^2\rangle_\mathrm{i} \langle \Hat{q}^2\rangle_\mathrm{det}$, 
with $\langle \Hat{O}\rangle_\mathrm{i} = \Tr_\mathrm{sys}(\Hat{O}\rho_\mathrm{i})$ average with the initial state of the system. 
We used the fact that, 
when the postselection in $E_\mathrm{f}$ fails, the system is postselected in the complementary state\footnote{In general, we could establish a multiple postselection through the following procedure: 
With probability $p_j$ a strong measurement of an observable $\Hat{S}_j$ out  of a set of preestablished arbitrary observables $\{\Hat{S}_1,\Hat{S}_2,\dots\}$ 
is made on the system after it interacted with the probe; the outcome $S$ is obtained; 
with an arbitrarily chosen probability $w(f|S,j)$, 
the outcome is given the label $f$. The system is thus postselected in the mixed state $\ensuremath{E_\mathrm{f}=\sum_{S,j} w(f|S,j)\, p_j |\Hat{S}_j:S\rangle \langle \Hat{S}_j:S|}$. 
Notice that $\sum_{E_\mathrm{f}} E_\mathrm{f}=1$ and that the $E_\mathrm{f}$ are not necessarily normalized to one $\Tr_\mathrm{sys}(E_\mathrm{f})=\sum_{S,j} w(f|S,j)\, p_j\neq 1$, in general.} 
 $1-E_\mathrm{f}$, so that  
$\Tr_\mathrm{sys}[(1-E_\mathrm{f})\Hat{A}^j\rho_\mathrm{i}\Hat{A}^k]=\langle\Hat{A}^{j+k}\rangle_\mathrm{i}-\Tr_\mathrm{sys}[E_\mathrm{f}\Hat{A}^j\rho_\mathrm{i}\Hat{A}^k]$, hence the parameters $\omega,\alpha,\beta$ become $\tilde{\omega}=1-\omega$, 
$\tilde{\alpha} = \langle\Hat{A}\rangle_\mathrm{i}-\alpha$, and $\tilde{\beta} = \langle\Hat{A}^2\rangle_\mathrm{i}-\beta$. 

The lack of normalization for the probabilities is a consequence of the approximate propagator 
$U_1 =1+i\lambda\Hat{A}\Hat{q}$ not being a unitary operator up to order $\lambda^2$. 
We need to normalize the expressions, 
\begin{align}
\P(o,E_\mathrm{f})\simeq&\ \frac{\omega \overline{\Pi_o} +\lambda \left\{ i\alpha \overline{\Pi_o\Hat{q}}+c.c.\right\}
+
\lambda^2 \beta\overline{\Hat{q}\Pi_o\Hat{q}}}{1+\lambda^2 \langle \Hat{A}^2\rangle_\mathrm{i} 
\overline{\Hat{q}^2}},
\label{joint}
\\
\P(E_\mathrm{f})\simeq&\ \frac{\omega +\left\{ i\lambda \alpha \overline{\Hat{q}} +c.c.\right\}+
\lambda^2 \beta \overline{\Hat{q}^2}}{1+\lambda^2 \langle \Hat{A}^2\rangle_\mathrm{i} \overline{\Hat{q}^2}}.
\label{marg}
\end{align}
 If one considers, as usually done in the context of weak measurement, 
the conditional probability $\mathcal{Q}(o)=\P(o,E_\mathrm{f})/\P(E_\mathrm{f})$ and its related averages, the normalization is ininfluent. 
Furthermore, the overlap $\omega$ can be simplified between numerator and denominator, and one can define the canonical weak value $A_w=\alpha/\omega$ and the positive real 
number $B_w=\beta/\omega$, reducing the parameters to two. While mathematician will shudder in disgust, 
$\omega$ may as well be 0, and the formulas still be valid, in the sense that in this limit $B_w$ is overwhelmingly large compared to $A_w$, in both the numerator 
$\P(o,E_\mathrm{f})$ and the denominator $\P(E_\mathrm{f})$. 
Another point in favor of this improved expansion is that when the preparation and postselection are orthogonal, i.e. $\omega\to 0$, while $\alpha$ tends to 0 as well, 
$\beta$ stays finite, excluding some trivial cases for which the probability of postselection is exactly null. 
For this reason, the expansion is robust for any preparation and postselection of the system. 

But what does a na\"{i}ve application of Taylor series, as learnt from Calculus, prescribes? 
Since we are including a second order term, according to the prescription, for consistency we should expand the propagator 
up to second-order, and retain terms like 
$\lambda^2 \Tr{\left[(E_\mathrm{f}\otimes|o\rangle\langle o|)\Hat{A}^2 \Hat{q}^2 (\rho_\mathrm{i}\otimes\rho_\mathrm{det})\right]}$.  
This was done in Ref.~\cite{Wu2011}, while Refs.~\cite{DiLorenzo2012a} and \cite{Kofman2012} stated that the reason to neglect 
these terms, which give rise to a complex number $C_w=\Tr_\mathrm{sys}[E_\mathrm{f}\Hat{A}^2\rho_\mathrm{i}]/\omega$, was that the second order correction becomes relevant only in the regime $|C_w|\ll B_w$. 
From the discussion above, it can be seen that the dropping of $C_w$ is further justified by the positive-definiteness of the probability. 
Thus, if we wanted to retain terms $\lambda^2\Hat{A}^2\Hat{q}^2$ in the propagator $\U$, we should retain the 
$\lambda^3$ and $\lambda^4$ terms in the probability that appear when multiplying the contributions from $\U$ and $\U^\dagger$. 

Finally, the conditional average of an arbitrary observable $\Hat{o}$ is obtained by replacing $\Pi_o$ with $\Hat{o}$.

%
\section{Proof of some inequalities}
We consider the complex vector space $\mathcal{L}$ formed by all linear operators $\Hat{X}$ acting on a Hilbert space $\mathcal{H}$. 
In particular, we fix two nonnegative linear operators $P_1$ and $P_2$. 
We define the scalar product
\begin{equation}
(\Hat{X},\Hat{Y}) = \Tr[P_2\Hat{X}P_1\Hat{Y}^\dagger]
\label{eq:scprod}
\end{equation}
It is immediate to verify that the definition \eqref{eq:scprod} satisfies the following properties: \\
(i) $(\Hat{X}+\Hat{Y},\Hat{Z})=(\Hat{X},\Hat{Z})+(\Hat{Y},\Hat{Z})$,\\
(ii) $(z\Hat{X},\Hat{Y})=z(\Hat{X},\Hat{Y})$, $\forall z\in \mathbb{C}$,\\
(iii) $(\Hat{Y},\Hat{X})=(\Hat{X},\Hat{Y})^*$,\\
(iv) $(\Hat{X},\Hat{X})\ge 0$.\\
However, depending on $P_1$ and $P_2$, there may be some nonnull operator $\Hat{X}\neq\Hat{0}$ that has zero length, 
$(\Hat{X},\Hat{X})=0$, i.e. in general the scalar product \eqref{eq:scprod} is positive semi-definite. Let us 
call $N_0$ the null space, $N_0=\{\Hat{X}\in\mathcal{L}:(\Hat{X},\Hat{X})=0\}$.

In the following, we characterize $N_0$ more precisely. We call $K_j$, $j=1,2$ the kernel of the operator $P_j$, namely 
$K_j=\{\psi\in\mathcal{H}:P_j\psi=0\}$. As is well known, $K_j$ are closed subspaces of $\mathcal{H}$.  
We call $C_j$ the orthogonal complement of $K_j$, so that $\mathcal{H}=C_j\oplus K_j$. 
In other words, $C_j$ is the subspace spanned by all the eigenvector of $P_j$ that do not correspond to a zero eigenvalue. 
Then it is easy to prove that $\Hat{X}\in N_0$ iff $\psi_2^\dagger \Hat{X}\psi_1=0$ for all $\psi_1\in C_1$ and $\psi_2\in C_2$. 
In practice, it is sufficient to verify this relation for the eigenstates $|f_1\rangle$ and $|f_2\rangle$ of $P_1$ and $P_2$, respectively, that generate the subspaces $C_1$ and $C_2$, i.e. that have nonzero eigenvalues. 

For the scalar product \eqref{eq:scprod}, the Cauchy-Schwarz inequality reads 
\begin{equation}
|(\Hat{X},\Hat{Y})|^2\le (\Hat{X},\Hat{X}) (\Hat{Y},\Hat{Y})
\label{eq:semics}
\end{equation}
with the equality sign only in one of the two cases:  
(1) $\Hat{X}$ or $\Hat{Y}$ belongs to $N_0$; 
(2) for some complex number $z$, $\Hat{X}-z\Hat{Y}\in N_0$.

Let us apply the inequality \eqref{eq:semics} to some cases of interest. 
\subsection{Justification of the inequality $|A_w|^2\le B_w$}
We specialize Eq.~\eqref{eq:semics} to the case $\mathcal{H}=\mathcal{H}_\mathrm{sys}$, and we put $P_2=E_\mathrm{f}$, $P_1=\rho_\mathrm{i}$. 
Then, for $\Hat{X}=\Hat{A}$, $\Hat{Y}=\id$, the Cauchy-Schwarz inequality reads 
\begin{equation}
|\alpha|^2\le \beta \omega, 
\label{eq:semics2}
\end{equation}
with $\alpha,\beta,\omega$ defined in Eq.~\eqref{eq:weaknorm}. 
If $\omega\neq 0$, after dividing by $\omega^2$, we get $|A_w|^2\le B_w$. 

The equality in Eq.~\eqref{eq:semics2} applies only (i) if $\id \in N_0$, i.e., the postselection is orthogonal to the preparation:  $\omega=\Tr[E_\mathrm{f}\rho_\mathrm{i}]=0$; 
or (ii) if $\Hat{A}\in N_0$, i.e., $\beta=0$; or yet (iii) if 
\begin{equation}
\Hat{A}=z \id + \Hat{X},
\label{eq:condeq}
\end{equation}
 with $\Hat{X}\in N_0$, in which case $A_w=z$ and $B_w=|z|^2$. 
We notice that, since $E_\mathrm{f}=\sum_{e_\mathrm{f}> 0} e_\mathrm{f} \Pi_\mathrm{f}$ and 
$\rho_\mathrm{i}=\sum_{w_\mathrm{i}> 0} w_\mathrm{i} \Pi_\mathrm{i}$, with $\Pi_\mathrm{f}=|f\rangle\langle f|$ and $\Pi_\mathrm{i}=|i\rangle\langle i|$ 
one-dimensional projection operators, 
then Eq.~\eqref{eq:condeq} implies that $\Tr[\Pi_\mathrm{f}\Hat{A}\Pi_\mathrm{i}]=\langle f|\Hat{A}|i\rangle\langle i|f\rangle=z|\langle f|i\rangle|^2$, 
$\forall f,i:e_\mathrm{f}>0,w_\mathrm{i}>0$. 

In particular, if $E_\mathrm{f}\propto |f\rangle\langle f|$ and $\rho_\mathrm{i}=|i\rangle\langle i|$ represent pure states, Eq.~\eqref{eq:semics2} holds with the equality sign, as can be seen by inspection. 
Indeed in this case Eq.~\eqref{eq:condeq} is trivially satisfied for any $\Hat{A}$: choose $z=\langle f|\Hat{A}|i\rangle/\langle f|i\rangle$; 
then, automatically, $\Hat{X}=\Hat{A}-z$ is a null-vector.

\subsection{Validity of the perturbative expansion}
After substituting Eq.~\eqref{eq:rhoaft} into Eq.~\eqref{eq:unnormp}
the Taylor expansion of the propagator yields 
\begin{align}
|M-M_1|&=
\left|\sum_{m,n}\nolimits'\frac{(-1)^n(i\lambda)^{m+n}}{m!n!} \overline{\Hat{q}^{n}\Hat{o}\Hat{q}^{m}}
\Tr_\mathrm{sys}[E_\mathrm{f}\Hat{A}^m \rho_\mathrm{i}\Hat{A}^n]\right|
\nonumber
\\
&\le \sum_{k=2}^\infty{\mathop{\sum\nolimits'}\limits_{n=0}^k}
\frac{|\lambda|^{k}}{k!}\binom{k}{n}
|\overline{\Hat{q}^{n} \Hat{o} \Hat{q}^{k-n}}| 
\left|\Tr_\mathrm{sys}[E_\mathrm{f}\Hat{A}^{k-n} \rho_\mathrm{i}\Hat{A}^n]\right|
\label{eq:ineq1}
\end{align}
where $\sum'$ in the first line means that the pairs $(m,n)\in\{(0,0), (0,1), (1,0), (1,1)\}$ are excluded. 
In the last line, we changed variables to $k=m+n$ and $n$. The primed sum means that if $k=2$ the value $n=0$ is excluded. 
After letting $\mathcal{H}=\mathcal{H}_\mathrm{sys}$, 
$P_2=E_\mathrm{f}$, $P_1=\rho_\mathrm{i}$, we get that the Cauchy-Schwarz inequality implies 
$|\Tr{(E_\mathrm{f}\Hat{A}^m\rho_\mathrm{i}\Hat{A}^n)}|=|(\Hat{A}^m,\Hat{A}^n)|\le (\Hat{A}^m,\Hat{A}^m)^{1/2} (\Hat{A}^n,\Hat{A}^n)^{1/2}$. 
Furthermore, 
\begin{align}
(\Hat{A}^m,\Hat{A}^m)=&\ \Tr{(E_\mathrm{f}\Hat{A}^m\rho_\mathrm{i}\Hat{A}^m)}=\sum_{f} e_\mathrm{f} \langle f|\Hat{A}^m\rho_\mathrm{i}\Hat{A}^m|f\rangle
\le \sum_\mathrm{f} e_\mathrm{f} \sum_{f'} \langle f'|\Hat{A}^m\rho_\mathrm{i}\Hat{A}^m|f'\rangle
\nonumber\\
=& \Tr[E_\mathrm{f}]\sum_A A^{2m}  
\langle A|\rho_\mathrm{i}|A\rangle
\le\ \Tr[E_\mathrm{f}] \max{\{A^2\}}^{m}.
\end{align}
Thus, Eq.~\eqref{eq:ineq1} yields 
\begin{align}
|M-M_1|\le& \Tr[E_\mathrm{f}] \sum_{k=2} \frac{|\lambda\max\{|A|\}|^{k}}{k!}
{\mathop{\sum\nolimits'}\limits_{n=0}^k}
\binom{k}{n}
|\overline{\Hat{q}^{n} \Hat{o} \Hat{q}^{k-n}}|.
\end{align}
In particular, for $\Hat{o}=\id$ we obtain the approximation for  Eq.~\eqref{eq:ppost1}
\begin{align}
|N-N_1|\le& \Tr[E_\mathrm{f}] \sum_{k=2} \frac{|\lambda\max\{|A|\}|^{k}}{k!}
{\mathop{\sum\nolimits'}\limits_{n=0}^k}
\binom{k}{n}
|\overline{ \Hat{q}^{k}}| \le 
\Tr[E_\mathrm{f}] \sum_{k=2} \frac{|2\lambda\max\{|A|\}|^{k}}{k!}
|\overline{ \Hat{q}^{k}}|.
\end{align}
If the inequality 
\begin{equation}
|2\lambda\max\{|A|\}|^{k}
|\overline{ \Hat{q}^{k}}|\le \delta^k,
\label{eq:cond0}
\end{equation}
holds, then 
\begin{align}
&|N-N_1|\le \sum_{k=2} \frac{\delta^k}{k!} \Tr[E_\mathrm{f}]= \Tr[E_\mathrm{f}](e^\delta-1-\delta)=\varepsilon,\ q.e.d.
\end{align}

Next, we note that 
$|\overline{\Hat{q}^{n} \Hat{o} \Hat{q}^{k-n}}|=|\Tr_\mathrm{det}\{[\Hat{q}^{n},\Hat{o}] \Hat{q}^{k-n}\rho_\mathrm{det}\}+
\Tr_\mathrm{det}\{\Hat{o}\Hat{q}^{k}\rho_\mathrm{det}\}|$. 
We conjecture that for the observables $\Hat{o}$ of interest 
$|\Tr_\mathrm{det}\{[\Hat{q}^{n},\Hat{o}] \Hat{q}^{k-n}\rho_\mathrm{det}\}|\lesssim
u|\Tr_\mathrm{det}\{\Hat{o}\Hat{q}^{k}\rho_\mathrm{det}\}|$, with $u$ a positive constant. 
Then, we apply the Cauchy-Schwarz inequality \eqref{eq:semics} 
with $\mathcal{H}=\mathcal{H}_\mathrm{det}$, 
$P_2=\id$, $P_1=\rho_\mathrm{det}$, so that 
$|\Tr_\mathrm{det}\{\Hat{o}\Hat{q}^{k}\rho_\mathrm{det}\}|\le \overline{\Hat{o}^2}^{1/2} \overline{\Hat{q}^{2k}}^{1/2}$
\begin{align}
|M-M_1|\le& (1+u) \Tr[E_\mathrm{f}] \overline{\Hat{o}^2}^{1/2}\sum_{k=2} \frac{|\lambda\max\{|A|\}|^{k}}{k!} \overline{\Hat{q}^{2k}}^{1/2}
{\mathop{\sum\nolimits'}\limits_{n=0}^k}
\binom{k}{n}
\nonumber\\
\le& 
(1+u) \Tr[E_\mathrm{f}] \overline{\Hat{o}^2}^{1/2}
\sum_{k=2} \frac{|2\lambda\max\{|A|\}|^{k}}{k!} \overline{\Hat{q}^{2k}}^{1/2}
.
\end{align}
If the inequality \eqref{eq:cond} holds, then 
$|M-M_1|\le (1+u) \overline{\Hat{o}^2}^{1/2} \varepsilon$. 
Notice how the condition \eqref{eq:cond} implies \eqref{eq:cond0}, so that \emph{a fortiori} $|N-N_1|\le \varepsilon$. 

%

\section{Vanishing average commutator and anticommutator} 
We shall treat the case $a=c=0$, and show that it can be obtained as a limiting case of the general result. 
Indeed, the planes \eqref{eq:planes} are parallel to the $xy$-plane, 
precisely $z=\langle\delta o\rangle/(s-\langle\delta o\rangle)$. For $\langle\delta o\rangle=0$ the plane is tangent to 
the paraboloid \eqref{eq:boundary} at the origin, so that the optimal weak value is $A_w^\mathrm{opt}=0$, and 
for  $\langle\delta o\rangle=s$ the plane is tangent to the paraboloid in the improper point at infinity, so that 
the optimal value for the maximum (or minimum if $s<0$) is $A_w=\infty$, i.e. the preparation and postselection must be orthogonal. Formally, this case can be obtained by taking the limit of Eqs.~\eqref{eq:result} and \eqref{eq:sol}. 
Indeed, the upper sign solution yields $\langle\delta o\rangle\to s$ and $x_0\to\infty, y_0\to\infty$, while the lower sign 
yields  $\langle\delta o\rangle\to 0$ and $x_0\to0, y_0\to0$.

\section*{Bibliography}
\bibliographystyle{num-names}

\end{document}